\documentclass[journal]{vgtc}                     

\graphicspath{{figures/}{pictures/}{images/}{./}} 
\usepackage{mathptmx} 
\usepackage{booktabs}
\usepackage{dirtytalk}
\usepackage{array}
\usepackage{soul}
\usepackage{multirow}

\usepackage{enumitem}
\setlist{nosep}

\usepackage[noorphans,vskip=2pt]{quoting}

\newcommand{\sys}[1]{\textsc{Texture}}
\MakeRobust{\sys}

\newcommand{\colname}[1]{{\texttt{#1}}} 

\preprinttext{}

\onlineid{1270}
\vgtccategory{Research}
\vgtcpapertype{analytics and decisions}

\title{Texture: Structured Exploration of Text Datasets}

\author{
  \authororcid{Will Epperson}{0000-0002-2745-4315},
  \authororcid{Arpit Mathur}{0000-0002-0776-6485},
  \authororcid{Adam Perer}{0000-0002-8369-3847},
  \authororcid{Dominik Moritz}{0000-0002-3110-1053}
}

\authorfooter{
All authors are with Carnegie Mellon University. 

Emails: willepp@cmu.edu, arpitmam@andrew.cmu.edu, adamperer@cmu.edu, domoritz@cmu.edu.
}

\abstract{%
Exploratory analysis of a text corpus is essential for assessing data quality and developing meaningful hypotheses.
Text analysis relies on understanding documents through structured attributes spanning various granularities of the documents such as words, phrases, sentences, topics, or clusters.
However, current text visualization tools typically adopt a fixed representation tailored to specific tasks or domains, requiring users to switch tools as their analytical goals change.
To address this limitation, we present \sys{}, a general-purpose interactive text exploration tool.
\sys{} introduces a configurable data schema for representing text documents enriched with descriptive attributes. 
These attributes can appear at arbitrary levels of granularity in the text and possibly have multiple values, including document-level attributes, multi-valued attributes (e.g., topics), fine-grained span-level attributes (e.g., words), and vector embeddings.
The system then combines existing interactive methods for text exploration into a single interface that provides attribute overview visualizations, supports cross-filtering attribute charts to explore subsets, uses embeddings for a dataset overview and similar instance search, and contextualizes filters in the actual documents.
We evaluated \sys{} through a two-part user study with 10 participants from varied domains who each analyzed their own dataset in a baseline session and then with \sys{}. 
\sys{} was able to represent all of the previously derived dataset attributes, enabled participants to more quickly iterate during their exploratory analysis, and discover new insights about their data.
Our findings contribute to the design of scalable, interactive, and flexible exploration systems that improve users' ability to make sense of text data.
}


\teaser{
  \centering
  \includegraphics[width=\linewidth, 
  alt={todo}]{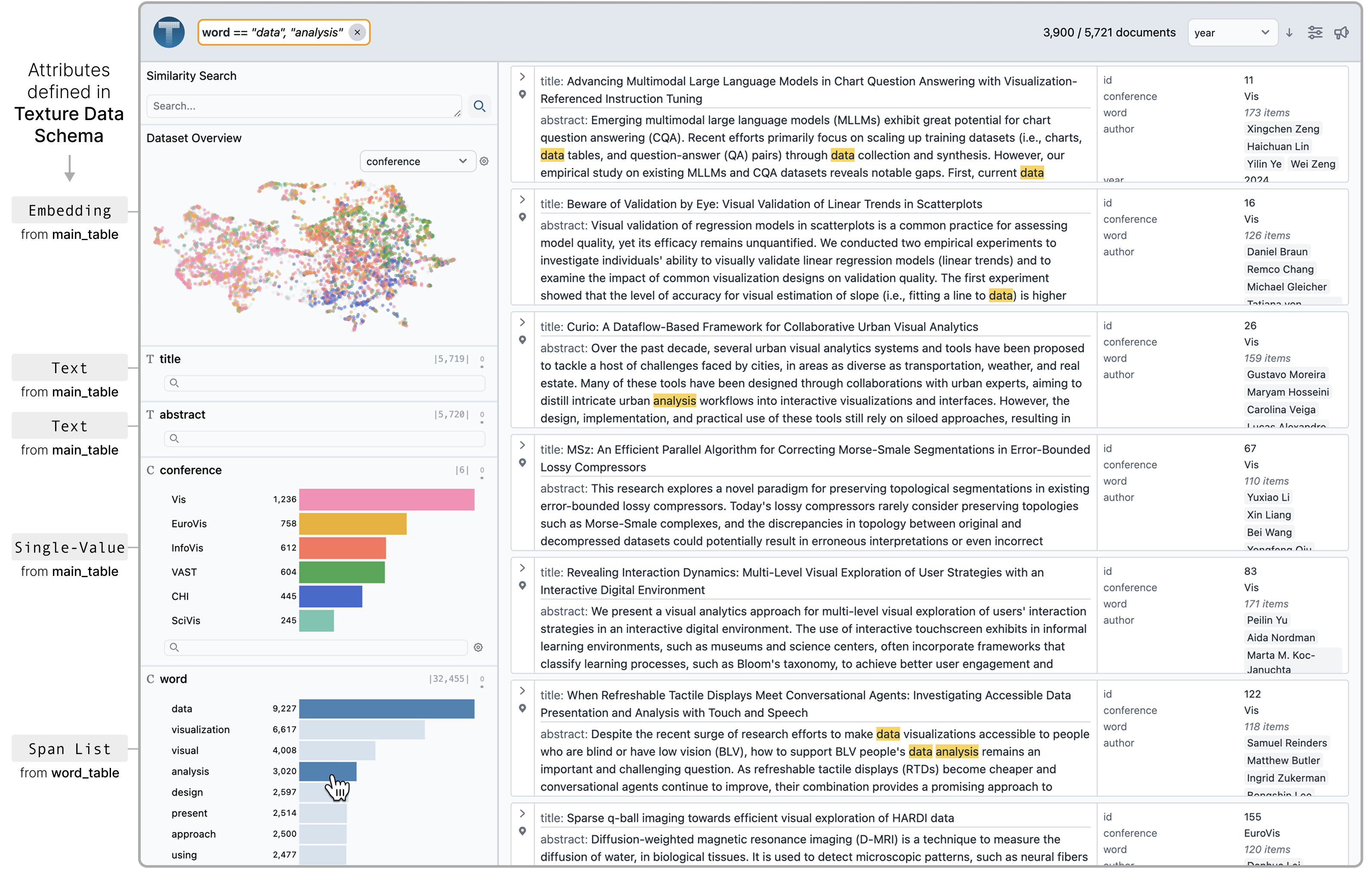}
  \caption{\sys{} helps users explore text datasets through structured descriptive attributes. 
  Its configurable data schema supports attributes at any level of granularity in the text, such as document-level attributes like the conference and embedding or word-level counts shown in this example analysis of an abstract corpus. The system organizes list attributes like words with multiple values per document into new tables and then joins tables to enable scalable filtering. \sys{} helps users explore their data through attribute overview visualizations, interactive filtering, embedding overview and search, and contextualizing filters in the document text. 
  }
  \label{fig: teaser}
}

\renewcommand{\manuscriptnotetxt}{}
\nocopyrightspace

\begin{document}

\firstsection{Introduction}
\maketitle

Understanding collections of text documents is a fundamental task across many disciplines. 
In the social sciences, researchers programmatically analyze trends from social media posts, news articles, and government reports~\cite{wanner2014state, gentzkow2019text, evans2016textforsocial}, while in NLP researchers use text datasets to train and evaluate new AI models~\cite{robertsonAngler2023, kahng2024llm}.
With the rise of large language models (LLMs), text data is becoming increasingly available and important in many high-stakes applications~\cite{reif2024automatic, Sambasivan2021EveryoneWT}.
Whether for research, model development, or decision-making, users have the need to quickly understand the text in their datasets, and evaluate the results of their analyses on the text~\cite{Tukey1980WeNB, reif2024automatic}.

Since manually reading all documents in a corpus is infeasible, analysts must transform text data into shorter, more interpretable representations~\cite{elazar2023WIMBD}.
Many techniques exist to derive these shorter representations, ranging from extracting words and phrases, to tagging documents with topics or classes~\cite{liu2018bridging}.
While some descriptive attributes like words or topics are generally useful across datasets, many meaningful representations are task and dataset specific. 
Over the course of an analysis, users often experiment with different representations of their data as their questions and goals evolve.

However, many prior text visualization tools focus on particular domains or tasks where the descriptive attributes are predetermined.
For example, domain-specific systems for social media content analysis~\cite{bosch2013scatterblogs2, alsakran2011streamit}, academic literature review~\cite{arpitVitality2022, an2024vitality2}, or NLP model evaluation~\cite{kahng2024llm, sievert2014ldavis} all extract predetermined representations.
Likewise, task-specific systems for word or topic understanding operate on a pre-determined level of granularity~\cite{reif2024automatic, fast2016empath}.
These specialized tools enable deep analysis of specific questions, however lack the flexibility needed to adapt to varied datasets or shifting analytical goals during exploratory analysis~\cite{ittoo2016text}. 
Nevertheless, many exploratory interactions---such as filtering data subsets, searching for textual patterns, or exploring document embeddings---are common across tasks and domains. 
This fragmented landscape motivates our central research question: \textit{How can we build a general-purpose, interactive text exploration tool?}

To address this question, we first introduce a configurable data schema for describing text data and associated descriptive attributes and then present an interactive interface for exploring a text dataset through its configured attributes.
Our schema describes attributes along two primary dimensions: their relationship to the document (i.e., Does the attribute correspond to the entire document or only particular spans?) and the cardinality of the attribute (i.e., Does the attribute have a single value per document or multiple values?).
Accordingly, our schema categorizes attributes into five distinct types: text documents, single-value attributes, list attributes, span list attributes, and vector embeddings. 
These types cover common text representations from different levels of granularity in the text like topics (a \textit{single-value} attribute that describes a document), authors (a \textit{list} attribute with multiple values per document), or frequent words and n-grams (\textit{span list} attributes that correspond to parts of the text).
Additionally, the proposed data schema prescribes how to split multi-valued list attributes into new relational tables to enable interactive filtering and exploration at scale.

We then designed \sys{}, an interactive text exploration tool based on this configurable data schema. 
\sys{} combines interactions from previous text analysis systems into a single general-purpose tool.
\sys{} is designed around three primary interactions: the ability to provide an \textbf{overview} of a text dataset through structured attribute visualizations, \textbf{filter} attribute visualizations for exploration, and link visualizations back to the raw documents to \textbf{contextualize} results.
\sys{} automatically generates interactive overview visualizations for each attribute in a dataset that support cross-filtering to explore insights across attributes.
Attribute visualizations link to a view of the raw documents, helping users contextualize summaries and filters in the actual document texts.
Finally, \sys{} incorporates embedding-based operations such as similarity search and dimensionality reduction overviews to enable fuzzy search and a dataset summary. 
\sys{} integrates seamlessly with Python-based data workflows, allowing analysts to programmatically define and manipulate descriptive attributes before exploring them interactively.
While prior text visualization systems contain subsets of these features, \sys{}'s novelty is the combination of features with a configurable data schema that enables exploration of many different types of text datasets.
\sys{} is open-sourced and available for download
\footnote{\url{https://github.com/cmudig/Texture}}. 

We evaluate \sys{} through a two part user study with 10 participants.
Each participant brought their own dataset they had previously analyzed to ensure higher validity of our study.
In the first session, they walked us through a baseline analysis of their data using their current workflow.
Participants used many different kinds of structured attributes to understand their text. 
However, they only inspected small samples of the text in their dataset and were slowed down by the need to manually define charts through code and their inability to link them back to the text.

In the next session, each participant analyzed their data using \sys{}. 
Across all 10 distinct datasets, our results show that \sys{} helped users effectively explore their data and uncover new insights. 
Participants quickly explored different hypotheses and iteratively developed meaningful analysis questions with \sys{}. 
Our system and study inform the design of configurable, general-purpose tools that better support users in exploring text datasets across diverse domains.
In summary, this paper contributes:
\begin{enumerate}
    \item A configurable data schema for describing text attributes from arbitrary levels of document granularity and cardinality.
    \item \sys{}, an interactive text exploration tool that helps users explore their data through overview visualizations, filtering, and contextualizing descriptive attributes.
    \item Results from a user study that show how \sys{} is expressive enough to analyze datasets from 10 different tasks/domains and helps each user effectively explore their data. 
\end{enumerate}

\section{Related Work}

Our paper builds on related work on interactive systems for exploratory data analysis and text visualization.

\subsection{The Need for Exploratory Text Analytics}

Text data inherently lacks the structure necessary for straightforward visual analysis and must be processed into meaningful, structured attributes for exploration~\cite{reif2024automatic}. 
Recent studies underscore the necessity of understanding both individual documents and their descriptive attributes within both local (individual document) and global (full corpus) contexts~\cite{documentResearch2025Gururaja}.
Even basic document metadata, such as the most frequent words, n-grams, or domains, can yield significant insights into a corpus and its quality~\cite{elazar2023WIMBD}. 
This form of corpus \textit{profiling} is essential both for analytical purposes as well as AI model training, where data quality is often undervalued and prior analyses have found that popular NLP benchmark datasets still contain poor quality or mislabeled data~\cite{Sambasivan2021EveryoneWT, swayamdipta2020datasetCart}.

Profiling dataset metadata attributes is a core aspect of EDA workflows where analysts summarize their data with visualizations to generate initial hypotheses for further analysis~\cite{Wongsuphasawat2019GoalsPA, tukeyEDA1977}.
Prior interactive visualization systems for tabular data facilitate EDA by automatically constructing visual representations of the data and enabling interaction to quickly filter and find meaningful data subsets~\cite{2016-voyager, kandelProfiler2012}.
Automatic visualization combined with interaction enables a fast feedback loop between asking questions about data and interpreting the results~\cite{Epperson2023DeadOA}.
While these approaches craft effective visual summaries of attributes, they lack the ability to contextualize attribute summaries in the actual documents needed for EDA over text.

\subsection{Exploring Text Through Structured Attributes Across Levels of Granularity}

Understanding a text corpus requires navigating information across multiple levels of granularity.
We review prior approaches for understanding documents across three common levels---words, documents, and the entire corpus---and their influence on the design of \sys{}.

\subsubsection{Word Frequency and Highlighting}
Words offer a natural starting point for summarizing text documents into a more concise representation.
Techniques such as word clouds and parallel tag clouds display frequently occurring words or sequences across a dataset, potentially faceted by structured attributes~\cite{viegas2008timelines, wordCloudFelix2018, collis2009paralleltc}.
Since the significance of individual words often depends on their surrounding context, systems frequently use highlighting to link words back to the surrounding context of the original document~\cite{don2007discovering, correll2011taggedLiterary, muralidharan2012supporting, strobelt2015HighlightGuidelines, oelke2011visual}.
Word-level tags like entities or part-of-speech are also common attributes for analysis.
For example, in Jigsaw users investigate the relationship between entities across documents in a dataset~\cite{stasko2007jigsaw}, and Automatic Histograms presents a technique for clustering entities into meaningful semantic groups using LLMs~\cite{reif2024automatic}.
Beyond individual words, other research summarizes text documents through short phrases that match a particular query or linguistic pattern~\cite{wu2020tempura, reiflinguisticDiv2023}.
\sys{} builds on prior word analysis techniques by abstracting these into a span list attribute where values correspond to any arbitrary segment of the text and individual occurrences are highlighted.

\subsubsection{Document-Level Attributes}
Beyond words, other research considers how to navigate a corpus through attributes assigned to each document, most commonly topics.
Topic modeling techniques represent each document as one or more topics, each summarized by characteristic words.
Latent Dirichlet Allocation (LDA) is a foundational method using this approach~\cite{blei2003lda}. 
Interactive systems help users explore a corpus through topics by showing the most frequent topics, filtering to documents that match a particular topic, and understand how topics evolve over time~\cite{alexander2014serendip, fast2016empath, liu2012tiara}. 
Domain-specific systems like PaperLens linking topic visualizations directly to a predefined set of structured attributes~\cite{lee2005paperlens}.
With word-based topic techniques, users often struggle to interpret abstract topics represented by lists of potentially ambiguous words. 
Recent methods like LLooM improve interpretability by framing topics through clear inclusion criteria generated by LLMs, such as direct questions (e.g., ``Does this text discuss sports?'')~\cite{lam2024conceptInduction}. 
\sys{} generalizes techniques for understanding topics to any document-level attribute by showing frequent values and helping users explore how document-level attributes correspond to common words or phrases.

\subsubsection{Corpus Overview Methods and Embedddings}
At the highest abstraction level, corpus visualization techniques enable exploratory analysis across entire document collections. 
Many of these systems incorporate structured metadata alongside raw text.
Leam provides an interface for applying basic text data transformations (e.g., text length extraction) then visualizing the results~\cite{griggs-leam-2021}. 
TextTile defines three fundamental analysis operations---filter, split, and summarize---allowing users to compare keyword summaries and attribute visualizations across facets of their dataset~\cite{Felix2017TextTileAI}. 
Systems like LLMComparator and Vitality help contextualize dataset attributes by linking document views and structured attribute visualizations for particular types of text data like LLM outputs or paper abstracts~\cite{kahng2024llm, arpitVitality2022}.
Our work extends these prior corpus-level approaches by considering a broader set of attribute types, then enables similar interactions for users to filter and explore subsets.

In addition to structured attribute summaries, document embeddings are commonly used to understand a corpus.
Representing documents with high dimensional embedding vectors has become a common technique to capture both the syntax and semantics of the document in a single representation~\cite{pennington2014glove, reimers-2019-sentence-bert, bert2019}.
These embeddings can then be used for many useful analysis tasks, such as projecting the embeddings down to two dimensions to enable an overview visualization using techniques like UMAP~\cite{mcinnes2018umap-software}, or finding nearby instances in the embedding space.
Prior interactive systems help users explore documents through embeddings.
For example, DocuCompass helps users link subsets of the embedding space to inline structured attribute overviews~\cite{heimerl2016docucompass}. 
Angler, an interactive visualization tool for prioritizing machine translation errors, uses UMAP projections to provide an overview of the data while overlaying the scatterplot with additional information, such as usage logs~\cite{robertsonAngler2023}. 
Similarly, WizMap tightly integrates a projection view with automatically-generated multi-resolution summaries to help users navigate through the large embedding space of documents ~\cite{wangWizMapScalableInteractive2023}. 
\sys{} adopts similar techniques to these prior approaches by including embeddings as a fundamental data type, used for a projection overview and finding similar instances.
Like prior work, \sys{} helps users make sense of embeddings through the structured attributes in the corpus.

\section{\sys{}: Interactive Exploratory Text Analysis}
\label{sec: system}

\sys{} builds on prior work for understanding text at different levels of granularity such as words, document-level attributes, and structured corpus metadata in two primary ways.
First, \sys{} presents a configurable data schema for representing different kinds of structured attributes.
Unlike prior systems that use a fixed set of attributes or only consider attribute types relevant for a particular analysis task, \sys{}'s attribute schema is highly configurable to analyze different types of descriptive attributes across datasets.
Second, \sys{} enables exploratory analysis through interactions designed around these descriptive attributes.
These interactions build on methods from prior work to enable users to explore their data through attribute overview visualizations, filter and compare filters on attributes, use document embeddings, and relate attributes back to the document text.
We first describe the data schema underlying \sys{}, then the interactive features of our system.

\subsection{Configurable Attribute Schema}
\label{sec: data schema}

\begin{table*}[ht]
    \centering
    \begin{tabular}{@{}lp{8cm}p{7cm}@{}}
        \toprule
        \textbf{Attribute Type} & \textbf{Description} & \textbf{Examples} \\
        \midrule
        \texttt{Text} & The text documents  & Paper abstracts, news articles, LLM prompts \\ 
        
        \texttt{Single-Value} & Descriptive attribute that only has one value per document and is \textit{numeric}, \textit{categorical}, or \textit{temporal}  & Publication date, document sentiment, topic\\ 
        
        \texttt{List} & Descriptive attribute with multiple values per document  & List of authors, keywords, or topics \\ 
        
        \texttt{Span List} & Descriptive attribute with multiple values per document where each value maps directly to a \textit{span} of the text  & Words, tokens, phrases, part of speech tags\\ 
        
        \texttt{Embedding} & High dimensional embedding + a 2D projection  & SBERT embeddings~\cite{reimers-2019-sentence-bert} with UMAP projection~\cite{mcinnes2018umap-software} \\ 
        \bottomrule
    \end{tabular}
    \caption{The semantic data schema used in \sys{} describes different kinds of attributes and their relationship to text documents.}
    \label{tab: data schema}
\end{table*}

\sys{} formalizes attribute definitions with a configurable data schema.
This schema supports common representations of text at different levels of granularity---including words, phrases, n-grams, topics, or document tags---along with arbitrary user-defined attributes.
Expressing attributes in the \sys{} data schema makes the system highly configurable for exploring different datasets.

The \sys{} data schema describes attributes according to their cardinality and how they correspond to the text. 
The five types in this schema are presented in \autoref{tab: data schema}.
The first type is \texttt{Text} data which describes the documents in a dataset.
Text data can be documents of any length from social media posts to books.
A single instance might have multiple text attributes. 
For example, paper titles and abstracts, Spanish to English translation pairs, or LLM prompts and responses. 

The other four types in the schema categorize descriptive attributes.
Single-value attributes have a single value per document and are numerical, categorical, or temporal.
These describe document-level attributes such as a single topic tag, publication date, or sentiment score.
This is distinct from multi-valued attributes that are lists.
Many common descriptive attributes are lists.
For example, each document can be divided into a list of words or tokens.
Likewise, a single document may have a list of topics or authors.
Any attribute with multiple values is a \texttt{List} attribute.
If each attribute item corresponds directly to a span of the text it is a \texttt{Span List} attribute.
Words, tokens, or phrases are all examples of span list attributes since they correspond to a specific span segment of the document. 
Word-level tags such as part of speech tags or entity tags can also be represented as span list attributes.
Span list attributes enable \sys{} to capture the hierarchical nature of text.
Documents can be broken down into arbitrary levels of granularity and then annotated with other possible data.
All that is necessary is to maintain which index the segment comes from.

The last type of our data schema are \texttt{Embedding} representations of the text.
Representing embeddings as an atomic type in \sys{} allows users to choose the model and projection method they prefer, and then analyze the results in the system.
Currently, \sys{} only supports a single embedding per instance but could be extended in the future to support multiple.

\sys{} does not automatically derive any descriptive attributes from the text; instead, it aims to decouple attribute derivation from representation and exploration.
Meaningful attributes are often dataset and task-specific and comprise a huge space of possible attributes.
\sys{} is un-opinionated about how attributes are derived, allowing users to write whatever code or use whichever models best fit their task.
Once they have attributes, they can describe them with the \sys{} data schema and then explore their results in the system.

\subsubsection{Representing Attributes in Relational Tables}
\label{sec: relational tables}

Our semantic schema prescribes how data should be formatted in physical tables to enable exploration in \sys{}.
A key goal of this representation is to support the different data types from the schema in \autoref{tab: data schema}, while also enabling scalable interactions through relational query languages like SQL.
While modern DataFrame libraries like Python Pandas~\cite{pandas-dev} permit arrays or tuples in columns, most relational databases require that tables are in first normal form and each attribute has only a single value~\cite{Codd1970RelationalModel}.
Normalizing attributes into different tables makes it easier to visualize attributes and perform interactive filtering at scale with SQL~\cite{heer2024mosaic}.

\begin{figure}[htb]
    \centering
    \includegraphics[width=.9\columnwidth, keepaspectratio,alt={TODO}]{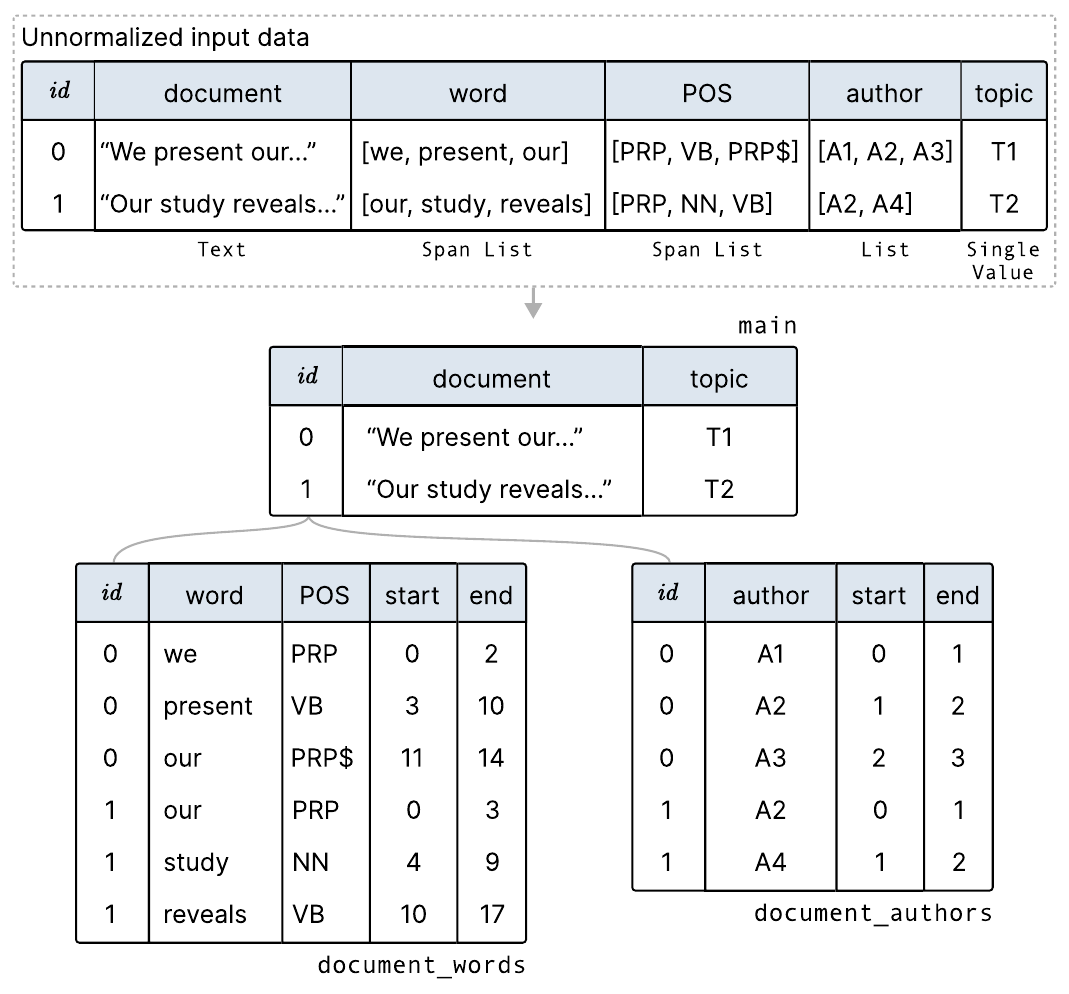}
    \caption{
    Following the \sys{} data schema requires placing list attributes into new tables that map back to the documents.
    }  
    \label{fig: normalized tables}
\end{figure}

Therefore \sys{} splits list attributes into new data tables that are linked back to the documents.
\autoref{fig: normalized tables} shows what this representation looks like for an example input dataset with different attribute types.
In this example, we have five attributes: one text, two span lists, one list, and one single value.
To achieve normalized tables, each of these list attributes is split into a new table while text and any single valued attributes stay in the same table.
In this example, each document and topic tag remain in the main table.
However the word, part of speech tags (POS), and authors are all lists which are parsed into new tables.

These tables enable representing hierarchy in the data. 
Each row in the main table has a unique id while the list attributes maintain a foreign key relationship to the ids.
Furthermore, for span list attributes, each table maintains the spans of the original document that the entry corresponds to.
These span indices are used for highlighting the text in the interactive system.
Regular list attributes maintain the array index.

\subsection{\textit{Overview:} Automatic Attribute Visualizations}
\label{sec: feat - attribute profile}
\begin{figure}[htb]
  \includegraphics[width=.6\linewidth]{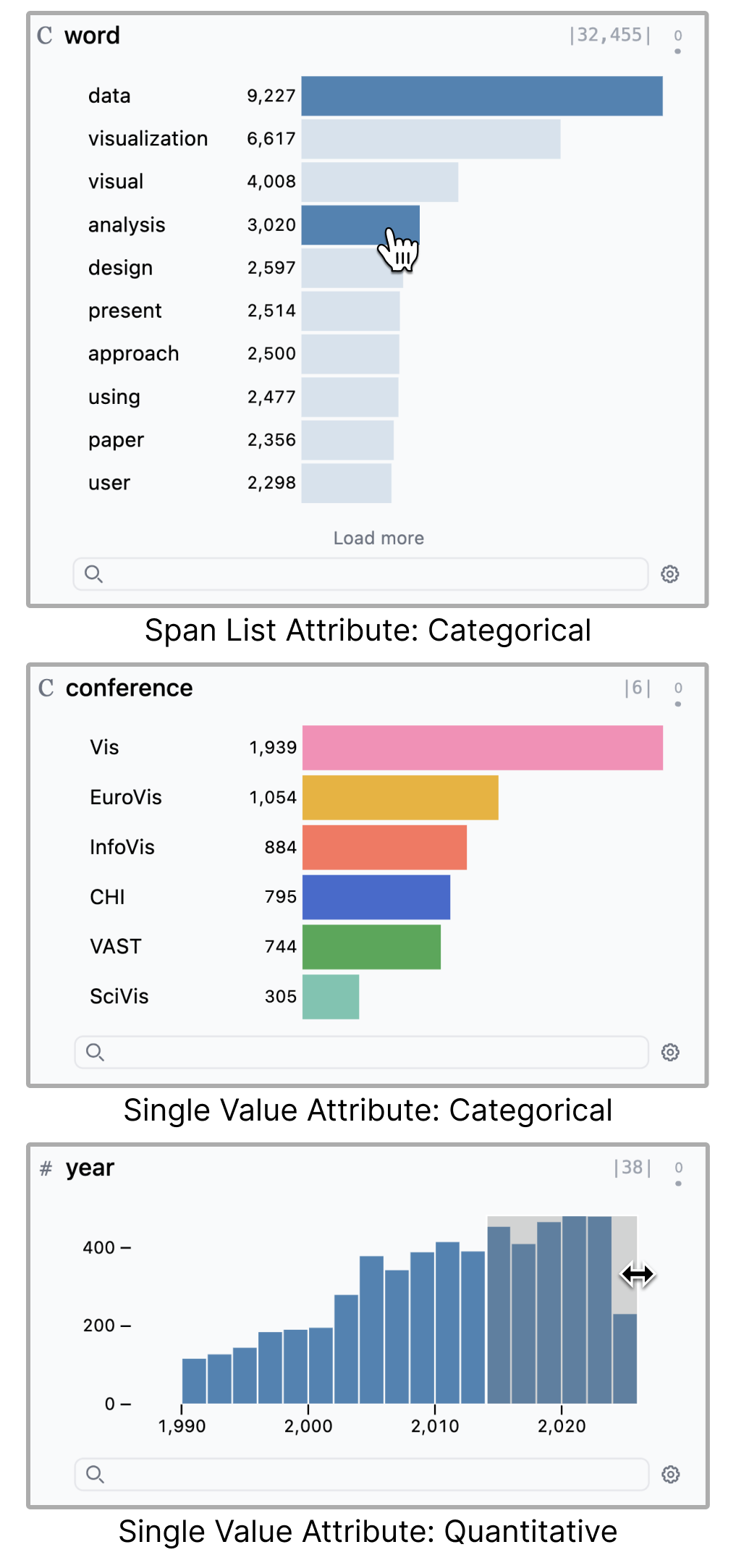}
  \centering
  \caption{
  All attributes are automatically visualized according to their data type (quantitative, categorical, or date) regardless of if they are lists or single-valued. Attribute visualizations support interactive cross-filtering and can color the projection overview.
  }
  \label{fig: attribute profile}
\end{figure}
Providing a dataset and schema to \sys{} enables immediate exploration in the UI.
\autoref{fig: teaser} shows \sys{} with the VisPubs dataset that was also used by one of the participants in our user study to understand visualization abstracts and is used throughout our examples~\cite{2024_preprint_vispubs}. 

The first component of this UI is the automatic visualization of each attribute into an interactive overview visualization.
These visualizations enable quick overviews of each attribute and support interactive filtering to enable users to explore documents in different subsets of the data.
Once formatted into separate tables, each attribute becomes tabular data.
We therefore designed the attribute overview visualizations similar to previous tabular data overview tools and text visualization systems that show structured attribute visualizations~\cite{Epperson2023DeadOA, kahng2024llm, arpitVitality2022}.

\sys{} automatically visualizes attributes according to their data type: text, quantitative, categorical, or temporal.
Example attribute visualizations are shown in \autoref{fig: attribute profile}.
Text columns are displayed in the main document view whereas categorical attributes are visualized.
Users can change the data schema to configure how attributes are displayed.
We produce a summary visualization for each attribute according to its data type:
\begin{itemize}
    \item \textbf{Quantitative} attributes are visualized as binned histograms.
    \item \textbf{Categorical} attributes are visualized as sorted bar charts of the counts of the 10 most frequent values (more on-demand)
    \item \textbf{Temporal} attributes show line charts of the count over time.
\end{itemize}

While each of the attributes might come from different tables, \sys{} shows the visualizations as a flat list to facilitate comparison between attributes while filtering.
This allows users to make corpus-level observations from list attributes.
For example, observing the most frequent words across the entire corpus or the most frequent author from lists of authors.

\begin{figure}[t]
  \includegraphics[width=.6\linewidth]{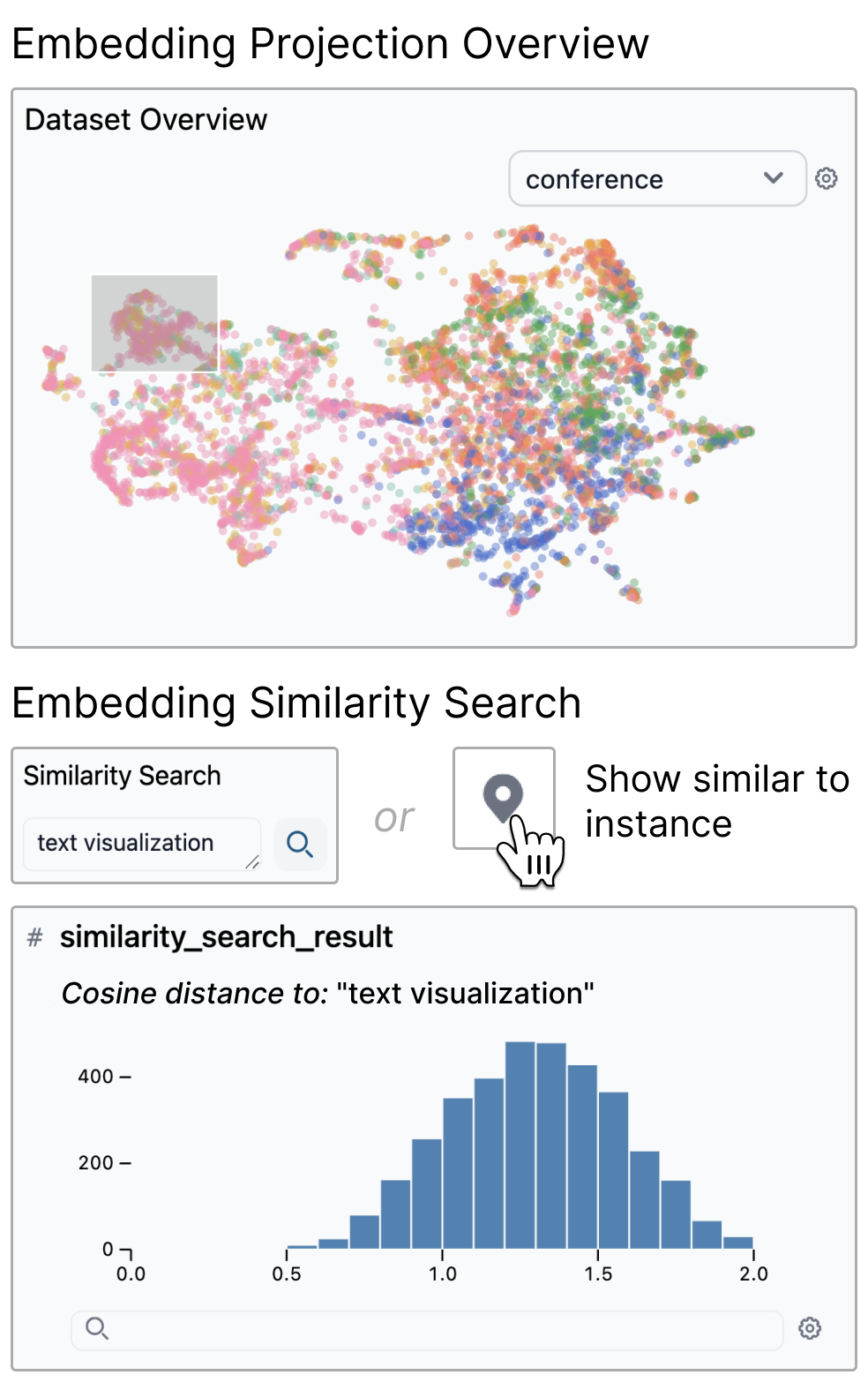}
  \centering
  \caption{
  Document embeddings enable a projection overview and similarity search.
  }
  \label{fig: embedding view}
\end{figure}

In addition to individual attribute visualizations, \sys{} provides a dataset overview plot through a scatterplot of 2D embedding projections.
This sort of dataset overview has become a common data overview technique for unstructured data like text in both research and commercial systems~\cite{wangWizMapScalableInteractive2023, nomicAI, heimerl2016docucompass}.
The projection view can also be colored by any of the single value categorical attributes in the dataset to help users plot clusters in the dataset.
For example, the embedding overview in \autoref{fig: embedding view} is colored by the conference attribute. 

\subsection{\textit{Filter:} Exploration Through Linked Visualizations}

After attribute visualizations provide an initial overview, \sys{} enables users to explore their data by filtering to different subsets.
Users can filter their data in several ways in the interface.
First, they can apply selections to visualizations to filter to a particular value or range of values.
As shown in \autoref{fig: attribute profile}, users can apply selections to categorical summary charts like the \colname{word} chart by selecting one or more bars, in this case the words \say{data} and \say{analysis}.
For quantitative or temporal charts, users can filter to a certain range with a brush.

Attribute visualizations are cross-linked, meaning that filters applied to one visualization filter the data in all the other visualizations, including the projection overview chart.
This enables insights beyond single attribute summaries by exploring interactions between attributes. 
For example, after filtering to certain values of \colname{word} in \autoref{fig: attribute profile}, a user can understand how these words are distributed across the \colname{conference} and \colname{year}.
Or a filter brushed across different years in the \colname{year} chart shows how the top words or conferences change across the years.
Filters can be applied to multiple charts at once, enabling questions about increasingly specific subsets of the data.

Users can also apply filters to the dataset overview chart to understand frequent attribute values for different regions of the projected embedding space.
Prior embedding visualization systems have directly labeled charts with frequent words or metadata values in different regions of the embedding space~\cite{wangWizMapScalableInteractive2023, heimerl2016docucompass}.
\sys{} does not label the embedding view directly, however enables a similar insight by cross-linking the embedding view to attribute visualization charts.
This allows users to interpret the embedding space through any of the attributes in their data, rather than just words.

Beyond filters on charts, \sys{} also supports search and embedding-based similarity search.
With the search bar, a user can search for a specific phrases in any of the text (or other) attributes and once again cross-filter the data.
Similarity search is supported in two ways, shown in \autoref{fig: embedding view}.
The first is open-ended similarity search where after a user enters a query, \sys{} calls a user-specified model to compute the embedding for the query, and calculates the cosine distance to each instance's embedding.
The second form is similar instance search where a user clicks the show similar button on an instance, then the system computes the cosine distance from this instance to all other in the corpus.
The result from either of these interactions is a new attribute that shows the similarity search result.
Like any other quantitative attribute, users can brush to different regions of similarity or sort their dataset by this value to find the most, or least, similar instances.

Since the values for \texttt{List} and \texttt{Span List} attributes in \sys{} are stored in different tables, \sys{} joins tables when relevant attributes are involved in a cross-filter selection.
This feature enables insights across attributes from different levels of a document hierarchy such as the aforementioned example about filters over a span list attribute \colname{word} with a document level attribute \colname{year}.
\autoref{appendix: filters across tables} discusses the query details on how \sys{} executes filters across joins.

\subsection{\textit{Contextualize:} Linking Attributes to Documents}
\label{sec: table view}

\begin{figure*}[htb]
  \includegraphics[width=.9\textwidth]{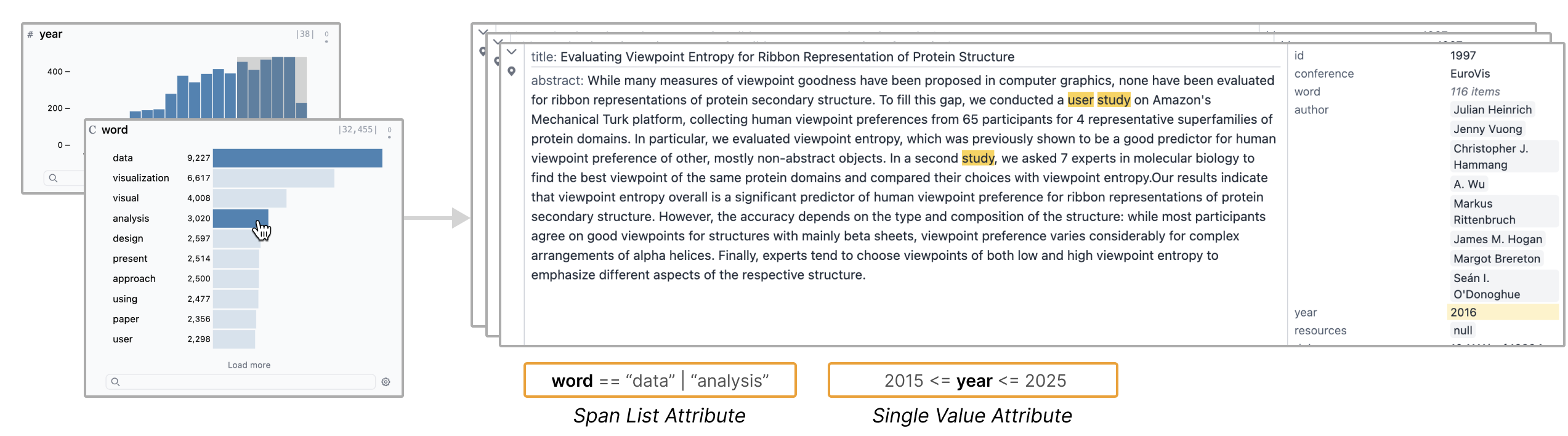}
  \centering
  \caption{%
    \sys{} helps users \textbf{contextualize} attribute filters in the actual documents by showing documents that match current filters and highlighting the spans of text for filtered span list attributes.
  }
  \label{fig: table view}
\end{figure*}

While attribute visualizations and filters help users make sense of the structured representations of their text, they also need to be able to contextualize attributes in the actual documents.
By design, a document table view occupies most of the screen in \sys{}.
This allows users to quickly inspect individual instances and their attributes.
The display table is scrollable with an entry for each instance that shows the text data and attributes.
The document view also allows sorting by each attribute.
Each instance shows the first five lines of text by default and can be toggled to show the entire document and other text attributes.

When filters are applied through attribute charts, the table view is filtered to the subset of documents that match the current filter.
Combined with attribute cross filtering, this enables users to quickly understand the results of filters both in terms of other attributes and individual documents.
For example, \autoref{fig: table view} shows how a user might apply filters to different attributes like year or words to inspect the subset of documents with the word \say{data} or \say{analysis} between the years 2015 and 2025.

Similar to prior word visualization systems~\cite{don2007discovering, alexander2014serendip}, \sys{} highlights spans in the text when filters are applied to span-list attributes.
The span indices maintained in the \sys{} data tables (discussed in \autoref{sec: relational tables}) ensure accurate highlights and disambiguate between substring matches.
For example, if a user filters to \texttt{word == "won"}, the span of this word is important to properly highlight only word matches.
This way \say{we \hl{won} the wonderful match} is highlighted and not \say{we \hl{won} the \hl{won}derful match}.

\subsection{Implementation \& Integration into Python Ecosystem}
The \sys{} interface runs in the browser and is built with a Svelte frontend and a Python FastAPI backend.
DuckDB SQL queries are generated by the frontend with queries coordinated by Mosaic~\cite{heer2024mosaic}.
DuckDB queries run in the backend.
\sys{} also uses a vector database, LanceDB, to store embeddings and calculate similarity distances between embedding vectors~\cite{lancedb}.

This Python-based implementation allows \sys{} to integrate into the Python data science ecosystem.
Python and computational notebook tools like Jupyter have become the most popular tools for programming with data~\cite{pythonOnGithub2024}.
Users can format their text data in Python, derive structured attributes to represent their data, then launch the \sys{} UI either from a Python script or computational notebook.

\section{User Study}
\label{sec: study design}

To evaluate how well \sys{} helps users performing different tasks understand their text data, we ran a user study.
We had two primary research questions for this study:
\begin{enumerate}
    \item How well can users working across different tasks represent their text datasets and descriptive attributes in the \sys{} data schema?
    \item Does \sys{} improve users' ability to understand their text, explore subsets, and find similar documents compared to their current workflows?
\end{enumerate}

\subsection{Procedure \& Methods}

To answer these questions, we designed a user study where users explored their own data in \sys{}.
Our study involved two separate one hour sessions: a baseline session and an experiment session.  

The baseline sessions were structured as a semi-structured think-aloud interview and live-coding session, where each participant walked us through an analysis on their own text dataset and explained their analysis goals. 
Participants first completed a background questionnaire on their text analysis experience and listed out their initial analysis questions about the data.
They then walked us through their analysis, showing us the kinds of visualizations and analysis steps they took to understand their dataset.
Afterwards, we conducted a semi-structured interview to probe about their experience.

In the next part of our procedure, each of these 10 participants also explored their data in \sys{} in a separate session.
Participants analyzed the same dataset with \sys{} as their initial exploration.
This session consisted of a semi-structured exploration, where participants were asked to think aloud while exploring their data analysis goals in \sys{}. 
Each participant sent their data to the research team before the session where we formatted the data into tables that fit the \sys{} data schema. 
We included the same attributes that participants explored in their initial sessions, derived words from the text attributes with the spans if words were discussed in the baseline interview, and added document embeddings for the text using the OpenAI text-embedding-3-small model~\cite{openAiEmbedding} if participants did not already provide them.
Each session lasted one hour, including a short demo of the features of \sys{} on a different dataset, 30 minutes for the participants to use the system, and an exit survey and interview.
In the survey, participants compared their experience using \sys{} to their baseline workflow.
They then provided open-ended responses elaborating on their ratings during a post-task interview.
The interview and survey questions asked in both sessions are provided in \autoref{appendix: user study questions}. 

Participants were compensated \$40 for the study, and the study protocol was approved by our institution's IRB. 
Both sessions were conducted over video call where audio and screen was recorded.
Interview transcripts and recordings of each participants' analysis actions were analyzed using thematic analysis to identify common patterns.

\subsection{Participant Details}

\begin{table*}[!ht]
    \centering
    \begin{tabular}{@{}llllrr@{}}
        \toprule
        \textbf{ID} & \textbf{Role} & \textbf{Data and Task} & \textbf{Attributes} & \textbf{Size} & \textbf{Med. \# Words} \\
        \midrule
        \textbf{P1} & VIS Researcher & Paper abstract corpus for literature review & 2 text, 10 descriptive & 5700 & 164   \\ 
        
        \textbf{P2} & NLP Researcher & LLM prompts and responses with constraints & 2 text, 3 descriptive & 1250 & 201  \\ 
        
        \textbf{P3} & NLP Researcher & AI agent responses & 2 text, 1 descriptive & 100 & 207 \\ 
        
        \textbf{P4} & NLP Researcher & LLM reasoning traces on benchmark & 3 text, 4 descriptive & 2600 & 1402 \\ 
        
        \textbf{P5} & HCI Researcher & Paper abstract corpus for literature review & 3 text, 4 descriptive & 3500 & 215\\ 
        
        \textbf{P6} & VIS Engineer & Song lyric corpus analysis & 3 text, 8 descriptive & 900 & 991\\ 
        
        \textbf{P7} & Humanities Researcher & Historical book corpus analysis & 1 text, 17 descriptive & 600 & 4218 \\ 
        
        \textbf{P8} & AI Engineer & LLM chatbot user queries and responses & 3 text, 5 descriptive & 16000 & 431 \\ 
        
        \textbf{P9} & AI Engineer & Reddit social media post analysis & 1 text, 10 descriptive & \textsuperscript{*}15000 & 18 \\ 
        
        \textbf{P10} & NLP Researcher & LLM fine-tuning dataset creation & 3 text, 8 descriptive & \textsuperscript{*}5000 & 1024\\ 
        \bottomrule
    \end{tabular}
    \caption{Participants in our study analyzed text data from a wide variety of domains and formats. We summarize the attributes in each dataset along with the size and median number of words in each document. \textbf{*}Indicates sample from a larger dataset.}
    \label{tab: participants}
\end{table*}

To be eligible for the study, participants had to have experience working with text data and programming in Python, and be able to bring a text dataset they had previously used in their work or research.
Participants were recruited through connections at our institution and advertisements on social media.

\autoref{tab: participants} details our study participants, their backgrounds, and high level analysis tasks. 
Participants had between 4 and 13 years of experience programming with Python (mean of 7.8 years), with self-reported expertise ranging from intermediate to expert.
Eight out of ten of our participants worked with text data daily or weekly, with the other two working with text data on a monthly basis.
Almost all of our participants brought datasets they had worked with in the past week, with only one participant bringing a dataset they had not worked with in over a year for confidentiality reasons.

Our participants primarily came from research backgrounds, analyzing research questions about text datasets or text-based AI models.
Some of their analysis goals were explicitly exploratory such as understanding themes in a corpus of song lyrics (P6) or getting a better sense of the literature in a field (P1, P5).
Other participants were seeking to better understand text datasets as part of a larger model building or evaluation task (e.g., P2, P3, P4, P10).

Participants analyzed datasets varying significantly in terms of content, structure, and scale. 
For instance, P7 analyzed the richest dataset, which included extensive descriptive attributes accumulated over multiple prior projects. 
Conversely, P3's dataset had minimal metadata, specifically a single categorical attribute identifying agent names for a multi-agent analysis. 
The datasets varied considerably in size, ranging from as few as 100 documents to as many as 16,000. 
Two of our participants (P9 and P10) analyzed samples of larger training corpora.

\section{Baseline Text Exploration Tasks and Challenges}
\label{sec: baseline findings}

For their baseline analysis, all participants were asked to show us how they explore their data through code and so spent most of the session stepping through their code-based EDA workflow.
Participants also used tools like the huggingface built in data viewer~\cite{huggingfaceDatasets2025} or spreadsheets to get an overview of their data.
Two participants also used custom-built tools for exploration.
P1 built a custom UI to inspect their data; P2 used the Zeno platform to inspect and filter their data~\cite{Cabrera2023ZenoAI}.
We describe some of the common analysis patterns and pain points from these baseline analysis sessions.

\subsection{Inspecting Small Sample of Documents}
A common task across domains involved inspecting a sample of individual data instances to generate hypotheses for further validation through exploration. 
8 of the 10 participants examined a few examples of text, often the first few rows of the dataset, to identify initial patterns within the data:
\begin{quoting}
\say{\textit{I think this is probably like the most common thing that I've done with almost every data set I've ever processed, which is you look at the first 30 rows and just read through these ones and you see a few things.}} - P8
\end{quoting}

However, with larger datasets, it becomes difficult to make sure these sample documents are a meaningful subset of the entire dataset.
Since hypotheses are generated by examining only a few instances, analysts might overlook important details in their documents.

\subsection{Exploration Through Structured Attributes}

Although participants were analyzing the \textit{text} in their data, each relied on different kinds of structured attributes to facilitate understanding their text. 
As shown in \autoref{tab: participants}, every participant used at least 1 structured descriptive attribute in their data, with up to 17.
The counts in this table represent the number of attributes that were in the data when passed to \sys{}, so includes both attributes in the original dataset as well as attributes derived over the course of the baseline session we observed.
For example, the abstract corpora analyzed by P1 and P5 each came with list attributes like the paper's authors, publication year, and conference.
The song lyric corpus analyzed by P6 included attributes like the song's artist, year, and popularity.
P7 described how their dataset and attributes were the combination of many prior projects where they had worked with the same collection of books.

Participants also derived new attributes to facilitate exploration.
Four participants derived the frequent words or n-grams in their dataset.
Three participants used the TextBlob package to quickly compute sentiment scores for their text~\cite{loria2018textblob}.
Six participants mentioned having previously used, or the desire to use, LLMs as a way to easily derive attributes for things that are hard to define in code.
For example, P10's dataset included three text attributes for LLM fine-tuning: a prompt, and two different model responses. 
They had previously used an LLM to extract the user-specified constraints on the output in each prompt and thus also had a list attribute for the constraints in each prompt.
 
Deriving useful attributes is an iterative process, with new attributes derived as new questions arise.
However, with each new attribute participants described how it can become hard to actually contextualize attribute summaries in the documents.
As P6 said when asked about what is currently the hardest part of their analysis:
\begin{quoting}
    \say{\textit{I think with text, it's converting words to numbers a lot. You're clustering, you're binning, you're counting, you're aggregating, \textbf{but you can lose the context very quickly.} So I would have to add a lot more code to just take one topic and then [look at] the lyrics for that topic. But if I could just select a topic and then say, `Show me five songs that identify strongest with that topic'...then I could read it if I wanted to}} -- P6
\end{quoting}

\subsection{Filtering Primarily Through Keyword Search}

Filtering is a fundamental analytical task in data analysis, and a similar pattern emerged during the baseline exploration. 
Participants applied filtering operations to both derived attributes (e.g., using word counts to exclude instances with short text) and metadata attributes (e.g., removing publications with low citation counts).

Keyword search was a common and easy form of filtering the text used to validate previously generated hypotheses. 
Participants filtered individual instances based on the presence of specific keywords. 
However, keyword search was not always sufficient to fully address the hypotheses, and participants expressed a need for other search approaches when exact matches are hard to find:
\begin{quoting}
    \say{\textit{I guess I can search for `predefined category', but it's fuzzy and kind of hard to use regular expressions here. I just need to understand how often people do this type of sentence or a variant of this sentence trying to constrain the classification results.}} -- P2
\end{quoting}

\subsection{Barriers to Using Embeddings}
The use of document embeddings for text similarity or overviews was not a common analysis task. 
Only one participant (P7) had previously computed embeddings for the dataset they brought.
Participants expressed interest in incorporating word embeddings into their workflows but cited the effort required use these embeddings for analysis as a barrier to adoption. 
For example, P4 described:
\begin{quoting}
    \say{\textit{Embeddings are great, but they're just hard to use...I think, honestly, [we have] TF-IDF and these simple things and word counts for a reason. So I just like quick and dirty.}}
\end{quoting}

\section{\sys{} Usage Results}

\begin{figure*}[ht]
  \includegraphics[width=\textwidth]{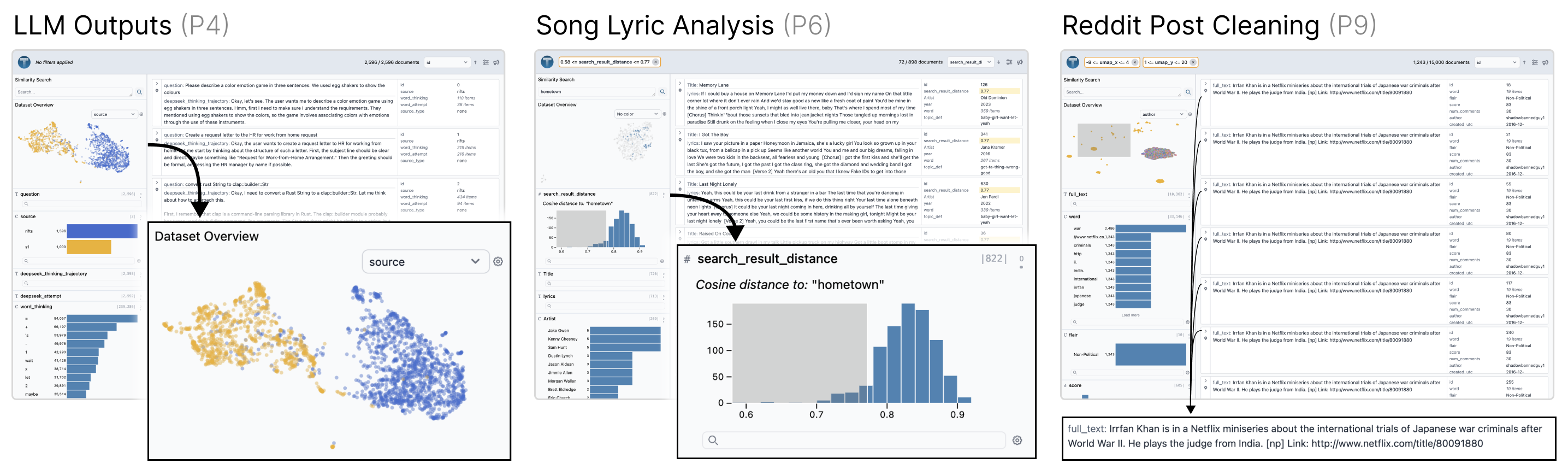}
  \centering
  \caption{
    Participants used \sys{} to explore a wide variety of datasets including LLM outputs, song lyrics, and Reddit posts.
  }
  \label{fig: demo gallery}
\end{figure*}

In this section, we discuss themes about participant's usage of \sys{} to explore their same dataset and how it compared to their baseline workflow.

\subsection{\sys{} is Expressive Enough for All Participant Attributes}

All of the diverse descriptive attributes from participants' baseline analysis could be represented in the \sys{} schema and analyzed with \sys{}.
Most attributes captured straightforward, document-level information and were modeled as single-value attributes.
However, eight out of ten participants also used list-like attributes with their datasets. 
Most of these were span list attributes for words from the different text attributes, but others included document-level lists such as authors (for P1 and P5) or lists of information extracted from the text (for P2). 
This ability to handle varied structured attributes underscores the flexibility and expressivity of \sys{}.

Exploring text data through the lens of structured attributes often inspired ideas for other useful attributes to add to the dataset.
During the exploration session, six participants added new attributes to their data and then explored the results by writing code in the Python notebook.
These were all simple attributes like the length of documents or number of words.
However, participants also mentioned how they might add other attributes like document topics, use LLMs to quickly derive new attributes for exploration, or find other datasets online to add to their current exploration given more time.

\subsection{\sys{} Makes Prior Analyses Faster and Enables New Types of Analyses}

With \sys{}, participants were able to perform the same analysis actions as in their baseline exploration, but more quickly and easily.
For example, every participant inspected attribute summary charts and applied multiple filters to attribute charts to explore subsets of their data.
This involved the keyword searches typical in the baseline sessions (also used by $7/10$ participants in \sys{}), along with filters over multiple attribute charts.

\begin{figure}[htb]
  \includegraphics[width=\linewidth]{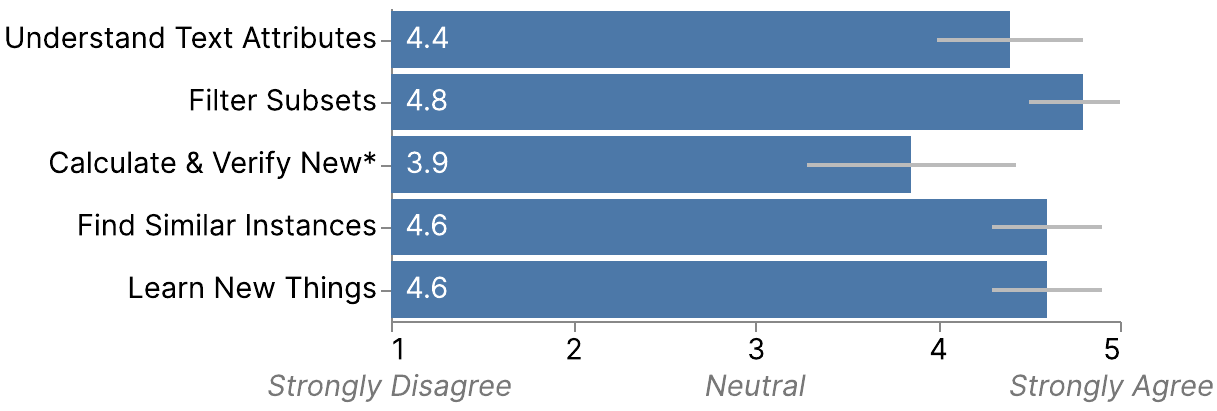}
  \centering
  \caption{
  Participants rated if Texture made it easier to perform certain actions relative to their baseline. Mean scores shown along with 95\% CI. \textbf{*}Calculate and verify new attribute reflects the seven participants who provided ratings for this aspect.
  }
  \label{fig: part 2 ratings}
\end{figure}

The automatic attribute visualizations in \sys{} sped up the analysis compared to creating charts by writing code in Python.
As P7 said, \say{This just gets me there so much faster}.
Overall, participants agreed that \sys{} made it easier to understand their text (mean rating $4.4/5$) and filter to subsets of their data (mean rating $4.8/5$, see \autoref{fig: part 2 ratings}).
In interviews, participants often attributed this improved ability in understanding their text to the fact that \sys{} encouraged them to read through more instances and made it easy to contextualize filters in the actual data:
\begin{quoting}
    \say{\textit{[Texture] makes me feel like I have more visibility into my data set...I'm immediately reading actual samples for like half the time, which I'm never doing in a notebook.}} -- P9
\end{quoting}

In addition to making prior workflows faster and easier, \sys{} also \textbf{enabled participants to explore their data in new ways}.
In our baseline sessions, only one participant had analyzed their dataset with embeddings for a projection overview and no participants leveraged similarity search. 
However, after we calculated embeddings for their data and uploaded them to \sys{}, participants found them to be a useful tool for understanding their data. 

All 10 of our participants used the embedding projection chart to get an overview of their dataset and explore different subsets of the embedding space. 
Even participants like P4, who mentioned in the baseline session that embeddings are too hard to use to be valuable, found immediate value from them in the tool when they noticed that the two main types of prompts in their data were clearly separated in the embedding plot (shown in \autoref{fig: demo gallery}): 
\begin{quoting}
    \say{\textit{That's great! The fact that it's already there. Literally all of what I did last meeting you just [see here]. This is exactly what I wanted to show}} -- P4
\end{quoting}

Once again, the ability to link filters across parts of the interface helped participants better understand their data.
For example, for P7 their core analysis task was to analyze historical books to find counterfeit publications in the late 1600s.
Using embeddings, their goal therefore was to find \say{unexpectedly similar documents}.
To do this, they colored the projection overview chart by different attributes like the political affiliation of a book and then looked for outliers--i.e., documents with similar embeddings but different political affiliations.
When they found an outlying book, they were able to then use \sys{} to read the text of this book and then compare the instance to others with similar attributes to better understand if it was actually counterfeit.

Whereas no participants used similarity search in their baseline explorations, the majority did while exploring their data in \sys{}.
Similar instance search was more common, used by seven participants whereas only two performed open ended text similarity searches.
Participants rated their ability to find similar instances as far easier in \sys{} than their baseline workflow ($4.6/5$).

\subsection{Exploring with \sys{} Leads to New Insights}

\sys{} provided flexibility for participants to explore data from multiple perspectives---transitioning between top-down and bottom-up modes of exploration.
Top down exploration involved looking at attribute summaries then contextualizing these attributes in the documents; bottom up involved finding an interesting document and then searching for the insight more broadly among documents with similar metadata or through similarity search.
These different analysis modes helped participants learn new things about their data they did not uncover in their prior analyses.
Overall, participants rated that they learned new things about their data with \sys{} as $4.6/5$ on average.

For example, P9 and P10 both found major data quality issues they were previously unaware of and did not discover in their baseline analyses.
P9 was working with a dataset collected from the social media platform Reddit that contained posts over a 4 year period.
They analyzed 15k posts in \sys{} out of their full 200k corpus. 
When initially reading through some of the instances to get a sense of the data, they noticed that one particular post seemed to re-occur.
Later, while inspecting different outlying clusters in the projection overview they noticed there was a large central cluster and many outlying clusters.
Inspecting these outlying clusters more, they noticed they were all the exact same post that comprised almost one third of their data (shown in \autoref{fig: demo gallery}).
They returned to their notebook, filtered out these duplicate points, then continued to analyze the remaining 10k posts in \sys{}.
They were previously unaware that this post existed so many times in their dataset.

Similarly, P10 identified cases of near repetition in their training prompts by inspecting outlier clusters in the embedding visualization.
By inspecting the data instances that corresponded to the different clusters, they noticed these were not exact duplicates but near duplicates where each prompt contained the same starting text and the end was slightly different.
These near duplicates comprised around 30\% of their overall dataset.
P10 emphasized the importance of dataset diversity for their training goals and noted this quality issue as something to address in future training.
When asked about how \sys{} helped them discover this new insight, they remarked:
\begin{quoting}
    \say{\textit{One huge value here is that AI people hate looking at data. Like I hate looking at data because it's just such a pain. But this makes it actually kind of fun to look at the data and definitely easy. So I'm pretty excited just using this now because I'm like, wow, I can actually see the data that I'm training models on.}} -- P10
\end{quoting}

Another theme across participant remarks was the speed with which they were able to explore and pivot between different analysis questions in \sys{}.
For example, P6 was analyzing a corpus of country song lyrics to see if they might contain useful signal around self-perceptions of rural identity.
Their analysis was inherently exploratory, where they were trying to better understand the data to see if it would even be useful for this task.
In their baseline exploration they had tried different analysis strategies like searching for keywords, looking at a song's sentiment, or basic keyword based topic models.
They began their analysis in \sys{} by inspecting different subsets of these same attributes, which then inspired them to do a similarity search for the term \say{hometown} (shown in \autoref{fig: demo gallery}).
From this, they found a particular song whose lyrics captured elements of home and country and then used this song to once again find similar instances.
These results contained many promising matches that they continued to explore for the rest of the analysis.
Reflecting on their analysis with \sys{}, they commented:
\begin{quoting}
    \say{\textit{I think the tool was really good at suggesting new questions, actually. Like, I had some, which I kind of talked to you about [in the first session], but it helped me narrow down the questions.}} -- P6
\end{quoting}


\subsection{Study Limitations}

Our study and findings are subject to potential limitations.
Our participants were recruited from a convenience sample and thus might not be representative of the larger population of text analysis practitioners.
In particular, our participants skewed towards computer science experts working in NLP.

Participants analyzed the same dataset in the first baseline session and the experiment session, which means additional insights could come simply from additional exposure.
However, since each participant analyzed a familiar dataset---previously explored in their own analysis or publication even before the baseline---we expect minimal learning effects from repeated exposure to the data. 

\section{Discussion}

\subsection{The Value of General-Purpose, Configurable Tools}
Many text visualization tools target specific analysis questions and methods from a particular domain or task. 
Such tools are highly valuable when participants are at the phase of analysis where they have decided on their research question and method.
However, with \sys{} we sought to explore a different goal of developing a general exploration tool across tasks as a way to support open-ended exploratory analysis.
With this approach, \sys{} \textbf{decouples the derivation of attributes from their exploration}.
This allows users to try different techniques---deriving common words, n-grams, topics, LLM-calculated topics, or another arbitrary method---and then explore the results in the same tool.
This approach is complementary to domain and task specific systems.
Exploratory tools for text like \sys{} facilitate fast first impressions of a data, verifying new attributes in a dataset, and, importantly, quickly determining if an analysis question is worthwhile before moving on to custom visual and analytical tools for a particular sub-task.
Several of the participants in our study commented on the value of this handoff, and how they would like to continue exploring a subset of the data or a question \textit{that they discovered in} \sys{}.

\subsection{EDA in the AI Coding Era}

In recent years, AI has begun to transform how people write code and interact with data through code~\cite{githubCopilot, productivityCopilot2022Ziegler}.
This surfaced in our study as well, where many participants used AI coding assistants to help write the code for their EDA or mentioned how they would like to process their text data further with LLMs.

What role do interactive data exploration systems serve in the era of AI coding?
Our results offer some clues.
Interactive systems can enable users to \textbf{discover interesting questions} about their data, which complements the ability of AI coding assistants to generate code to answer a specific question.
Much of the challenge and value of exploratory analytics from the start has been to develop enough intuition about the data to know what questions are worth asking~\cite{Tukey1980WeNB}.
Despite using AI coding assistants in their baseline sessions, our study participants still reported how \sys{} enabled them to quickly explore their data and come up with better analysis questions than they were able to do on their own.

These approaches---interactive systems and AI assistance---need not be in conflict but rather can complement one another.
Historically, this has been explored through paradigms like mixed initiative interfaces where a system can both allow user interactions and suggest actions~\cite{horvitz1999principlesMI}.
Exploratory text analysis tools like \sys{} can enable these sort of interactions at the workflow level, where an AI coding assistant suggests analysis code and then interactive tools help users to quickly inspect the results.
This combines the flexibility of writing code with the speed of interactive interfaces for data exploration, an observation noted in prior interactive data programming tools~\cite{mageKery2020}.

\subsection{Towards Rapid and Flexible Attribute Derivation}

A key element of data exploration, particularly for text data, is the availability of meaningful attributes to summarize and filter the data.
In our study, many participants had previously spent considerable time constructing such attributes with off-the-shelf libraries or custom methods for things like topics or sentiment.
Many participants expressed the desire for easier ways to derive task-specific attributes catered to their current analysis.
Particularly with the power of general-purpose LLMs for processing text, several participants expressed how it should be straightforward to derive custom attributes by transforming their data with a LLM.

We envision two different ways where interactive systems can better support such an LLM-assisted new attribute derivation workflow: (1) interactively deriving and verifying new attributes, and (2) suggesting interesting questions for derivation.
By integrating the ability to derive new attributes directly into systems like \sys{}, users could quickly derive an attribute, verify the results, and then use this attribute for further analysis.
Future research might investigate how to design such interactions to make this feedback loop as fast as possible, and how analysts use them in practice.
The second, and perhaps more difficult extension, would be to automatically generate potential questions for derivation, in the spirit of mixed-initiative interactions.
Recent systems like Automatic Histograms have begun to move in this direction by automatically grouping text entities into dataset overviews using LLMs~\cite{reif2024automatic}.
\sys{}'s data model offers a potential starting point for thinking about a broader set of attributes that LLMs can derive that correspond to the entire document or portions of the text.
Such methods would continue to increase the speed with which analysts can quickly structure their text data and then use interactive systems to inspect subsets and find interesting insights in their data.

\section{Conclusion}
In conclusion, we present the design, implementation, and evaluation of \sys{}---a configurable and general-purpose interactive system for exploratory text analytics.
\sys{} is built on top of a configurable data schema for representing different kinds of descriptive attributes alongside text.
We demonstrated the power of this data model to represent 10 different real-world datasets.
Participants in our study using \sys{} for analysis were able to more easily understand their data from both top-down and bottom-up analysis paths, using attribute overviews to summarize their dataset then drilling down to find interesting subsets and instances that inspired future analysis directions and questions.
Our system and study lay the groundwork for developing expressive and effective interactive systems for exploratory text analytics. 

\acknowledgments{%
Many thanks to Katelyn Morrison, Venkatesh Sivaraman, Sherry Wu, Niki Kittur, and Gagan Bansal for their feedback over the course of this project.
}

\bibliographystyle{abbrv-doi-hyperref}

\bibliography{refs}

\clearpage

\appendix 

\section{User Study Questions}
\label{appendix: user study questions}

Participants were asked the following questions during the surveys and semi-structured interviews in our study.

\subsection{Background questions}
Survey questions:
\begin{enumerate}
    \item How would your rate your experience with Python?
    \item How many years have you been using Python?
    \item How often do you work with text data?
    \item When was the last time you worked with this dataset in particular?
    \item What are the analysis questions you were trying to answer with this data?
\end{enumerate}

\subsection{Session 1: baseline exploration questions}

Discussion questions:
\begin{enumerate}
    \item How did you understand what is in the dataset? Can you show us which code or visualizations you used?
    \item Were all these attributes already present in the data? If not, how did you derive them?
    \item How did you verify or explore the results after adding a new attribute?
    \item What kind of filtering or subset analysis did you do of this dataset?
    \item Do you look at similar instances or use embeddings as part of your exploration / how?
    \item What was the hardest part about this task and working with this text data for you?
    \item If you had a tool to help you understand the text data, what do you wish it could do?
\end{enumerate}

\subsection{Session 2: \sys{} usage questions}

We asked the following Likert-scale questions in a survey immediately after the task:
\begin{enumerate}
    \item I found it easier to understand the text attributes in my dataset with Texture than the last session (5 point Likert scale)
    \item I found it easier to filter to subsets of my dataset with Texture than the last session (5 point Likert scale)
    \item I found it easier to calculate a new attribute and verify its correctness with Texture than the last session (5 point Likert scale)
    \item I found it easier to find similar instances in my dataset with Texture than the last session (5 point Likert scale)
    \item I learned new things about my dataset while exploring it with Texture (5 point Likert scale)
\end{enumerate}

Discussion questions:
\begin{enumerate}
    \item Talk me through your rating on how \sys{} impacted your understanding of your text attributes.
    \item Talk me through your rating on how \sys{} impacted your ability to filter to subsets.
    \item Talk me through your rating on how \sys{} impacted your ability to add a new attribute and verify.
    \item Talk me through your rating on how \sys{} impacted your ability to make sense of similar instances.
    \item Talk me through your rating on how you learned new things about your dataset with \sys{}.
    \item What other features would better support your workflow?
\end{enumerate}

\section{Technical Details: Filters Across Multiple Tables}
\label{appendix: filters across tables}

In \sys{}, attributes may come from different tables.
Therefore we must be able to filter across joined tables to enable interactive cross-filtering.

In Mosaic, filters are represented in SQL queries as where clauses.
For instance, when a user clicks on a document-level attribute like topic = \say{T1} in an attribute visualization, this is translated to a query like:

\medskip
\noindent\texttt{%
SELECT * \\
FROM main \\
WHERE topic = 'T1';
}
\medskip

However, when a user applies a filter to an attribute from another table, these tables must be joined together to make sure the data is properly filtered.
For example, if we apply a filter to \colname{document\_words} and are visualizing an attribute from another table like \colname{main} then we have to join the two tables:

\medskip
\noindent\texttt{%
SELECT * \\
FROM main JOIN document\_words USING (id) \\
WHERE word = 'study';
}
\medskip

\sys{} tracks the source table of each attribute and automatically joins tables as needed to support filtering across multiple tables.

The example queries here are not aggregated. Since most charts show aggregates, typically the join would be followed by an aggregation. To support fast cross-filtering over these aggregates, the underlying Mosaic system pre-computes aggregates. For details on the query construction and optimizations, refer to the Mosaic paper~\cite{heer2024mosaic}.

\end{document}